\documentclass[twocolumn,final]{revtex4-1}
\usepackage[utf8]{inputenc}
\usepackage{amsmath}
\usepackage{amsfonts}
\usepackage{amssymb}
\usepackage{graphicx}
\usepackage[ngerman,british]{babel}
\usepackage{color}
\let\oldsqrt\sqrt
\def\sqrt{\mathpalette\DHLhksqrt}
\def\DHLhksqrt#1#2{%
\setbox0=\hbox{$#1\oldsqrt{#2\,}$}\dimen0=\ht0
\advance\dimen0-0.2\ht0
\setbox2=\hbox{\vrule height\ht0 depth -\dimen0}%
{\box0\lower0.4pt\box2}}

\begin{document}

\title{The Mechanism behind Erosive Bursts in Porous Media}

\author{R. Jäger} 
\email{jaegerr@ethz.ch} 
\affiliation{  ETH
  Z\"urich, Computational Physics for Engineering Materials, Institute
  for Building Materials, Wolfgang-Pauli-Strasse 27, HIT, CH-8093 Z\"urich
  (Switzerland)}

\author{M. Mendoza} 
\email{mmendoza@ethz.ch} 
\affiliation{  ETH
  Z\"urich, Computational Physics for Engineering Materials, Institute
  for Building Materials, Wolfgang-Pauli-Strasse 27, HIT, CH-8093 Z\"urich
  (Switzerland)}
  
\author{H. J. Herrmann}
\email{hjherrmann@ethz.ch} 
\affiliation{  ETH
  Z\"urich, Computational Physics for Engineering Materials, Institute
  for Building Materials, Wolfgang-Pauli-Strasse 27, HIT, CH-8093 Z\"urich
  (Switzerland)}

\begin{abstract}
Erosion and deposition during flow through porous media can lead to large erosive bursts that manifest as jumps in permeability and pressure loss. Here we reveal that the cause of these bursts is the re-opening of clogged pores when the pressure difference between two opposite sites of the pore surpasses a certain threshold. We perform numerical simulations of flow through porous media and compare our predictions to experimental results, recovering with excellent agreement shape and power-law distribution of pressure loss jumps, and the behavior of the permeability jumps as function of particle concentration. Furthermore, we find that erosive bursts only occur for pressure gradient thresholds within the range of two critical values, independent on how the flow is driven. Our findings provide a better understanding of sudden sand production in oil wells and breakthrough in filtration.     
%
%
%
%
%
%
%
\end{abstract}

\maketitle
Erosion and deposition give rise to a plurality of applications and phenomena: Filtration in industrial processes, internal erosion in dams \cite{bonelli2013erosion}, and braided rivers \cite{schumm1972experimental}, to name but a few. Some of these phenomena have grave economical and monetary impact such as the performance reduction in water treatment filters or the sand production in oil wells \cite{zhou2016sand}. While it is well known that changes in flow conditions or internal structure can lead to erosion in filters \cite{NAG154,Han20091171,Kim2012433, Alem2015}, only recently Bianchi et al. \cite{filippo} conducted experiments to study the appearance of erosive bursts in porous media, finding that the resulting jumps in pressure loss follow a power-law distribution. In these experiments  suspensions of deionized water carrying $50 \mu m$ quartz particles are pushed with a peristaltic pump through a filter made of 1 mm glass beads measuring simultaneously pressure drop, flux and particle concentration.

Similar behavior has been observed in the past in different experiments. Fluctuations in the permeability were found in acidic flow experiments through porous rocks, where dissolution and precipitation alter the porous structure \citep{AIC:AIC690350713,WRCR:WRCR9267,WRCR:WRCR9499}. Sahimi et al. \cite{Sahimi2000} also measured such fluctuations in fractured carbonate oil reservoirs. They proposed that either dissolution or hydrodynamic forces are responsible for the increase in permeability. Nevertheless, the internal mechanism responsible for the power-law distribution of erosive bursts is still unclear. Here, we address this issue by analyzing the erosive process at pore scale employing advanced numerical techniques.

Numerical models have been already used extensively to study erosion and deposition in porous media. For instance, Mahadevan et al. \cite{mahadevan2012flow} introduced a model that shows how flow induces channelization in porous media. Kudrolli et al. \cite{PhysRevLett.117.028001} found that the porous medium evolves into a configuration that minimizes erosion. Bonelli et al. \cite{bonellisuffusion} introduced a model to describe suffusion. 
\begin{figure}
\includegraphics[trim = 2cm 2cm 2cm 2cm, width=0.8\columnwidth]{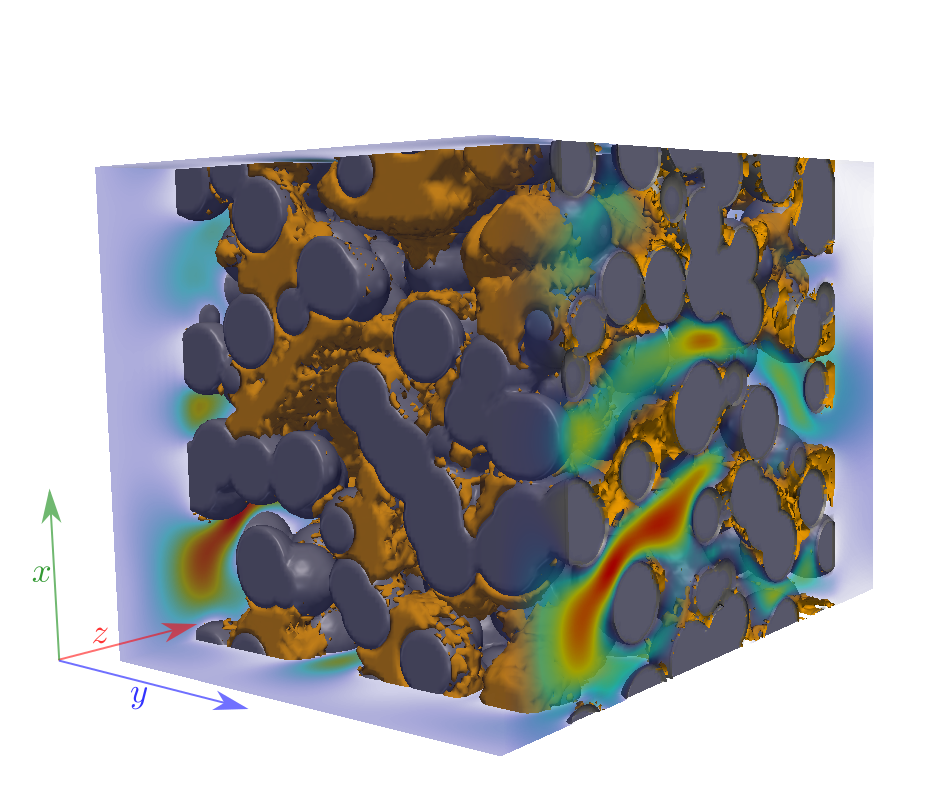}
\caption{Setup for simulations imposing constant flux in the $z$-direction, grey is the non-erodible initial porous matrix, brown is deposited matter and other colors indicate the magnitude of the flow velocity $|\vec{u}|$ of the fluid, where blue denotes low, and red high speeds.}
\label{fig:setup}
\end{figure}

Yamamoto et al. \cite{yamamoto} solved numerically the convection-diffusion equation using the lattice Boltzmann Method (LBM) simulating the flow of soot suspended in exhaust gas. Soot particles in contact with the filters' surface are deposited with a given probability $P_D$. Using an X-ray tomography (CT) scan of a diesel particulate filter they reproduced the deposition of soot and the increased pressure drop inside the filter. However, to the best of our knowledge, there are no models able to describe erosive bursts in porous media.

Based on the work of Yamamoto et al. \cite{yamamoto}, we proposed an extended model \cite{PhysRevE.95.013110} that allowed us to study erosion due to shear force, showing that static channels can form but no re-opening of clogged channels was observed. This model was not able to reproduce the erosive bursts found in Bianchi's experiments \cite{filippo}, and therefore, another mechanism for erosion must be acting. We introduce here an additional erosive mechanism where deposited matter is eroded if a critical pressure gradient is exceeded, which can lead to the unclogging of jammed channels. This is motivated by the fact that high hydraulic pressure leads to fluidization in granular matter, which after a decompaction phase can develop finger-like channels \cite{PhysRevE.78.051302}. In this letter, we show that this new ingredient is able to reproduce erosive bursts found in deep filtration, including its power-law behavior. Furthermore we study the conditions under which erosive bursts occur and find phases where bursts are predominant and others where they are not observed. This happens whether the flow through the porous medium features a constant flux or a constant pressure loss. The evolution of the porous structure, consisting of the filter and the deposited matter, can be compared to the evolution of braided rivers \cite{schumm1972experimental}, while some channels clog or change shape, others will unclog through erosion.


In our model, the flow is governed by the Navier-Stokes equations and the suspended particles are described by their particle concentration following the convection-diffusion equation \cite{PhysRevE.95.013110}:
\begin{equation}
\frac{\partial C}{\partial t} + \nabla \cdot(C\vec{u}) = \nabla \cdot(D\nabla C),
\end{equation}
where $C$ is the particle concentration, $\vec{u}$ is the fluid velocity and $D$ is the diffusion coefficient. Here we will consider $D$ to be equal to the kinematic viscosity and hence the Schmidt number is $Sc = \nu / D = 1$. While that may be much lower than for common fluids, we have found that the qualitative behavior does not change for higher Schmidt numbers ($Sc \approx 100$, see Supplemental Material \ref{sec:app}). To solve the Navier-Stokes and the convection-diffusion equations we employ a lattice Boltzmann method \cite{Pan2006898,talon2012assessment,yamamoto,PhysRevE.95.013110}. The fluid solver yields flow properties, such as the velocity $\vec{u}$ or hydraulic pressure $P$ on lattice sites.

Shear erosion occurs when the shear stress exerted by the fluid overcomes the cohesive strength of solid matter \cite{bonellisuffusion}. Our previous model \cite{PhysRevE.95.013110} only incorporated such a shear dependent erosion and was motivated by the work of Yamamoto et al. \cite{yamamoto}. Shear erosion is proportional to the difference between the wall shear stress $\tau_w$ and a threshold for shear erosion $\tau_c$ \cite{HYP:HYP10351}: $\dot{m} = - \kappa_{er}(\tau_w -\tau_c)$, where $\dot{m}$ is the eroded mass per area and time, and the threshold $\tau_c$ and the coefficient $\kappa_{er}$ depend on the specific composition of the solid matter and the fluid. We introduce here another erosive mechanism besides the wall shear stress. This erosion occurs when the hydraulic pressure gradient acting on deposited matter exceeds a certain threshold $|P_a - P_b| / L_{ab} > \sigma_c$, where $P_a$ and $P_b$ are the hydraulic pressure acting on deposited matter between two opposite points $a$ and $b$, $L_{ab}$ is the distance between $a$ and $b$, and $\sigma_c$ is a material specific positive constant. The hydraulic pressure $P_a = \sum_{\alpha,\beta}\pi^{\alpha\beta} n_\alpha n_\beta$ is calculated from the momentum flux tensor $\pi ^{\alpha \beta} = P \delta^{\alpha\beta} + \rho u^\alpha u ^\beta - \sigma^{\alpha\beta}$ \cite{PhysRevE.65.041203} and the unitary vector $\vec{n}$ normal to the solid surface at point $a$ (details are shown in the Supplemental Material \ref{sec:app}). 

While the erosion due to wall shear stress is a continuous process grinding away matter at the surface, the hydraulic pressure induces erosion that suddenly removes large pieces of deposits from the static matrix. We model this detachment by dissolving the deposited matter layer by layer into the fluid flowing through the porous matrix while locally increasing the concentration of suspended particles. After the erosion of the first layer the pressure on the new surface is similar to the pressure before, but the distance is shorter, thus fulfilling automatically for the next layer the condition for erosion. The erosion continues through the deposited material until a channel opens and the hydraulic pressure equalizes. Solid matter is represented by a mass index ($0 \leq m \leq 1$) on the same lattice which we use to identify the interface between solid and fluid. The hydraulic pressure is measured for all interface cells and the maximum gradient determined. When the criterion for erosion is fulfilled deposited matter in the cells is dissolved.

For channel flow the wall shear stress is proportional to the flux, hence increasing the flux or decreasing the wall shear threshold has the same effect. We write the critical wall shear threshold in dimensionless units by dividing by a characteristic shear, the same can be done for the threshold of the hydraulic pressure induced erosion $\sigma_c$:

\begin{equation}
\mathcal{T}_c = \frac{\tau_c l^*}{\rho \nu u^*} , \qquad
\mathcal{F}_c = \frac{\sigma_c l^*}{\rho u^{*2}},
\end{equation}

where $l^*$ is a characteristic length, $\rho$ the fluid density, $\nu$ the kinematic viscosity and $u^*$ the characteristic speed. The characteristic speed is flux divided by the cross area $\Omega$ of the porous system, the density $\rho$ and the porosity $\epsilon$: $u^* = \Phi / \Omega \rho \epsilon$.

To test our model, we perform numerical simulations, with a setup similar to filtration experiments \cite{Han20091171,Alem2015,filippo}. The filter consists of a porous matrix of non-erodible matter (see Fig. \ref{fig:setup}). This porous matrix is static and presents an upper bound for the permeability. The simulation box is a rectangular lattice of size $100\times100\times140$ ($xyz$), where inlet ($z<20$) and outlet ($z>120$) are kept free from solid matter and the length of the porous matrix is $L=100$. The porous matrix is built from randomly placed spheres with diameter $l^*=14.1$ lattice cells. They can overlap and are placed in the simulation box until a porosity of $\epsilon \sim 0.5$ is reached. The outlet ($z=1.4L$) has open boundary conditions and transverse directions ($x,y=\{0,L\}$) have periodic boundaries.
Just as in filtration experiments a fluid carrying particles flows through the filter and some particles can deposit onto the solid matrix. We impose either a constant flux or constant pressure loss through the porous medium and at the inlet the particle concentration is kept constant.
\begin{figure}
\includegraphics[width=\columnwidth]{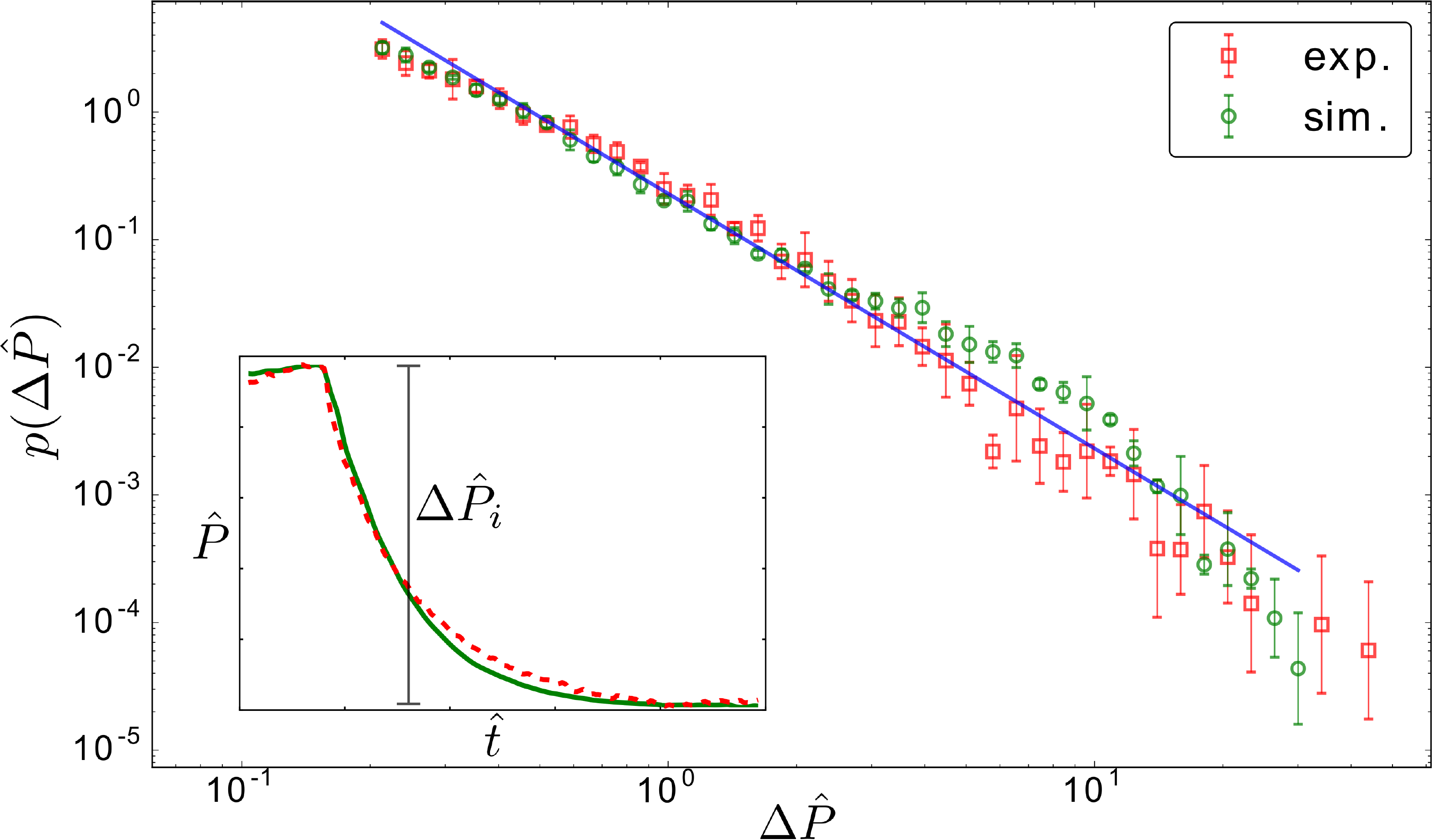}
\caption{Histograms of the size of pressure jumps of experimental \cite{filippo} and simulation data. The jumps are rescaled by the average jump height ($\Delta \hat{P}_i = \Delta P_i/ \Delta \bar{P}$). The slope of the histogram indicated by the blue line was found to be $-2.0 \pm 0.1$. The inset shows an example of a typical jump in the experimental (red dashed line) and the simulation data (green line), the $x$-, and $y$-axis are dimensionless pressure ($\hat{P}=P/\Delta P$) and time ($\hat{t}=t/\Delta t$, $\Delta t$ is the duration of the jump). }
\label{fig:jumpstat}
\end{figure} 

First we study the evolution of the filter when a constant flux with Reynolds number $Re=u^*l^* /\nu = 0.17$ is imposed. Each simulation starts with a clean porous matrix which does not contain erodible matter. The fluid carrying suspended particles (constant concentration at the inlet $C_0 = 10\%$ of total volume) is then pushed through the porous structure and deposition of particles leads to a decrease in permeability. While the permeability decreases, the macroscopic fluid pressure measured between the in- and outlet ($P=P_{\text{in}}-P_{\text{out}}$) of the porous structure keeps increasing. The hydraulic pressure gradient can increase even more on a local scale, which also increases the force acting on deposited material. When this force is high enough deposited matter detaches suddenly, i.e. much faster than deposition and shear erosion alter the surface. We indeed find that the macroscopic pressure during an erosive burst jumps and its decay (see inset in Fig. \ref{fig:jumpstat}) is very similar to the one reported in experiments \cite{filippo}. Furthermore we find that the size distribution of these jumps follows a power-law with an exponent $\alpha_{\text{sim}}=2.0\pm0.1$ (see Fig. \ref{fig:jumpstat}), which agrees within error bars with the exponent found by experiments $\alpha_{\text{exp}} = 1.88 \pm 0.09$ (\cite{filippo}). This slope does not depend on the inlet flow speed and hence the Reynolds number (see Supplemental Material \ref{sec:app}). Thus our simulation adequately describes the erosive bursts occurring in deep filtration experiments, and the erosive mechanism due to hydraulic pressure is apparently responsible for the experimentally observed bursts. 

\begin{figure}
\includegraphics[width=\columnwidth]{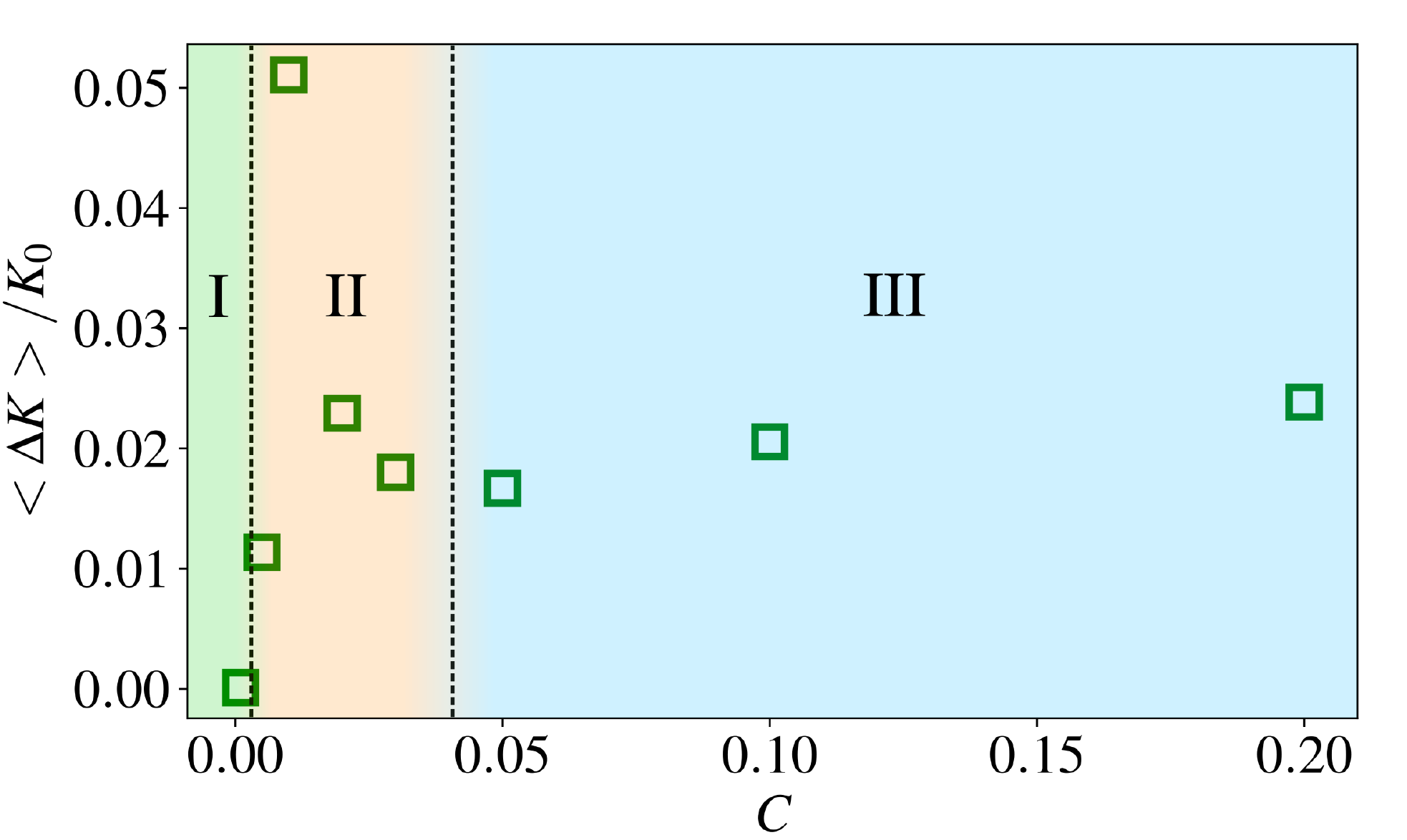}
\caption{The average relative jump in permeability $<\Delta K> /K_0$, where $K_0$ is the initial permeability. In the first regime (I) there are no jumps in permeability, in the second (II) there are jumps, and in the third (III) there are jumps but the porous medium eventually clogs.
}
\label{fig:conc}
\end{figure}

Bianchi et al. \cite{filippo} found that there is a minimum concentration required for erosive bursts to appear. This is also confirmed by our simulations. In Fig. \ref{fig:conc}, we observe that the bursts start to appear for concentrations around $C \sim 1\%$. Furthermore, in the experiments the presence of three different phases was reported: for low concentration, erosive bursts do not appear and no clogging is present; for intermediate particle concentrations, there are erosive bursts but no subsequent clogging; and for high concentrations, erosive bursts occur and the porous medium eventually clogs. We find the same qualitative behavior as in the experiments (see Fig.~\ref{fig:conc}), where the constant flux is only imposed up to a maximum pressure loss (see Supplemental Material \ref{sec:app} for details). Furthermore, we also found that the frequency of the bursts increases linearly with the concentration (see Supplemental Material \ref{sec:app}).


\begin{figure}
\includegraphics[width=\columnwidth]{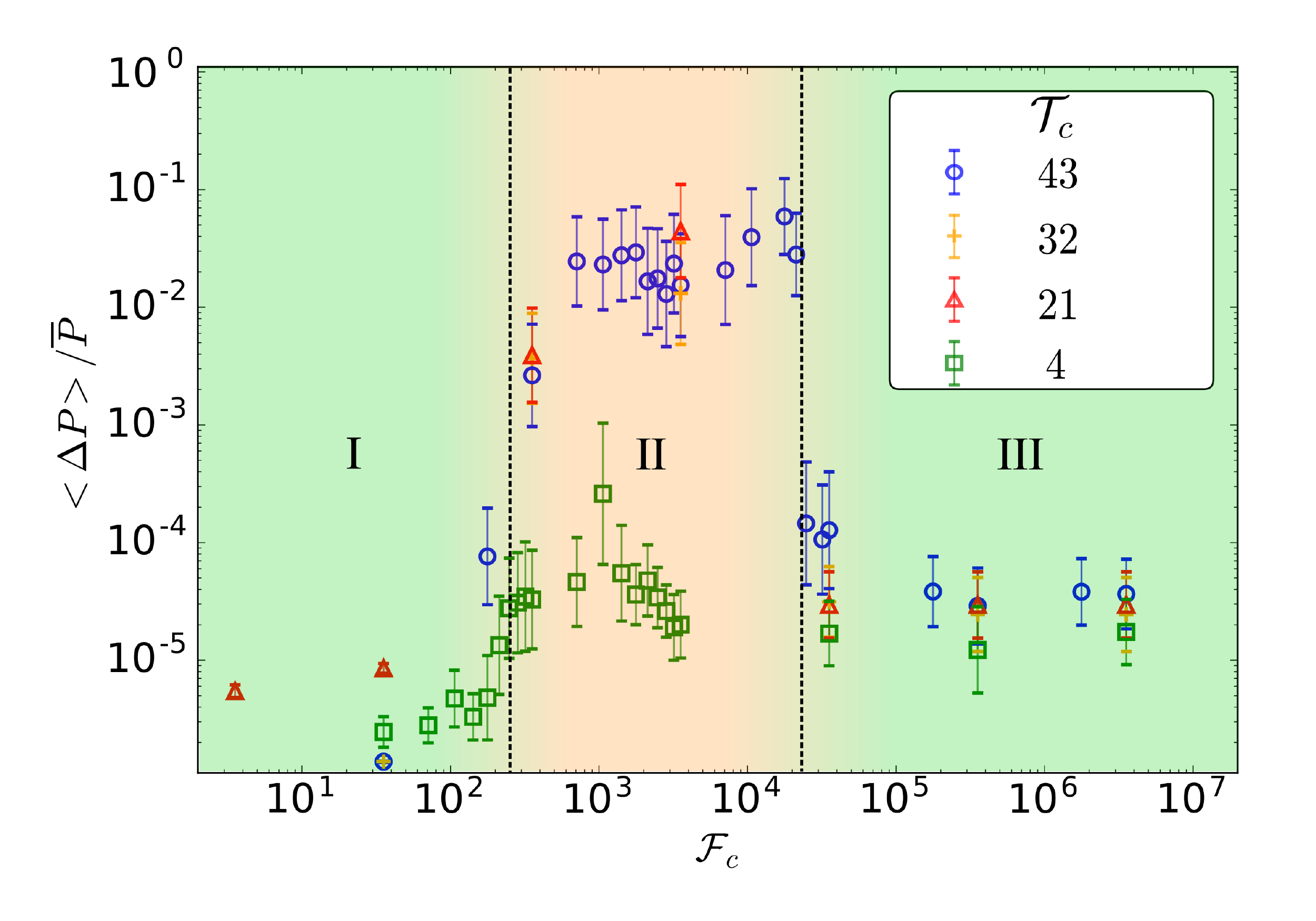}
\caption{The average jump size $< \Delta P>$ normalized by the average pressure $\bar{P}$ is shown for different values of $\mathcal{F}_c$ and $\mathcal{T}_c$. While for low values (I) and high values of $\mathcal{F}_c$ (III) there are no jumps, for an intermediate range (II) there are substantial jumps.
}
\label{fig:peakstudy}
\end{figure}

Additionally we found that depending on $\mathcal{T}_c$ jumps occur only within a certain range of $\mathcal{F}_c$ (see Fig. \ref{fig:peakstudy}). When $\mathcal{F}_c$ is below this range there is no clogging of pores and hence no erosive bursts. For high values of $\mathcal{F}_c$ there is no erosion due to hydraulic pressure. Thus we find three phases separated by two critical values of $\mathcal{F}_c$.

\begin{figure}
\includegraphics[width=\columnwidth]{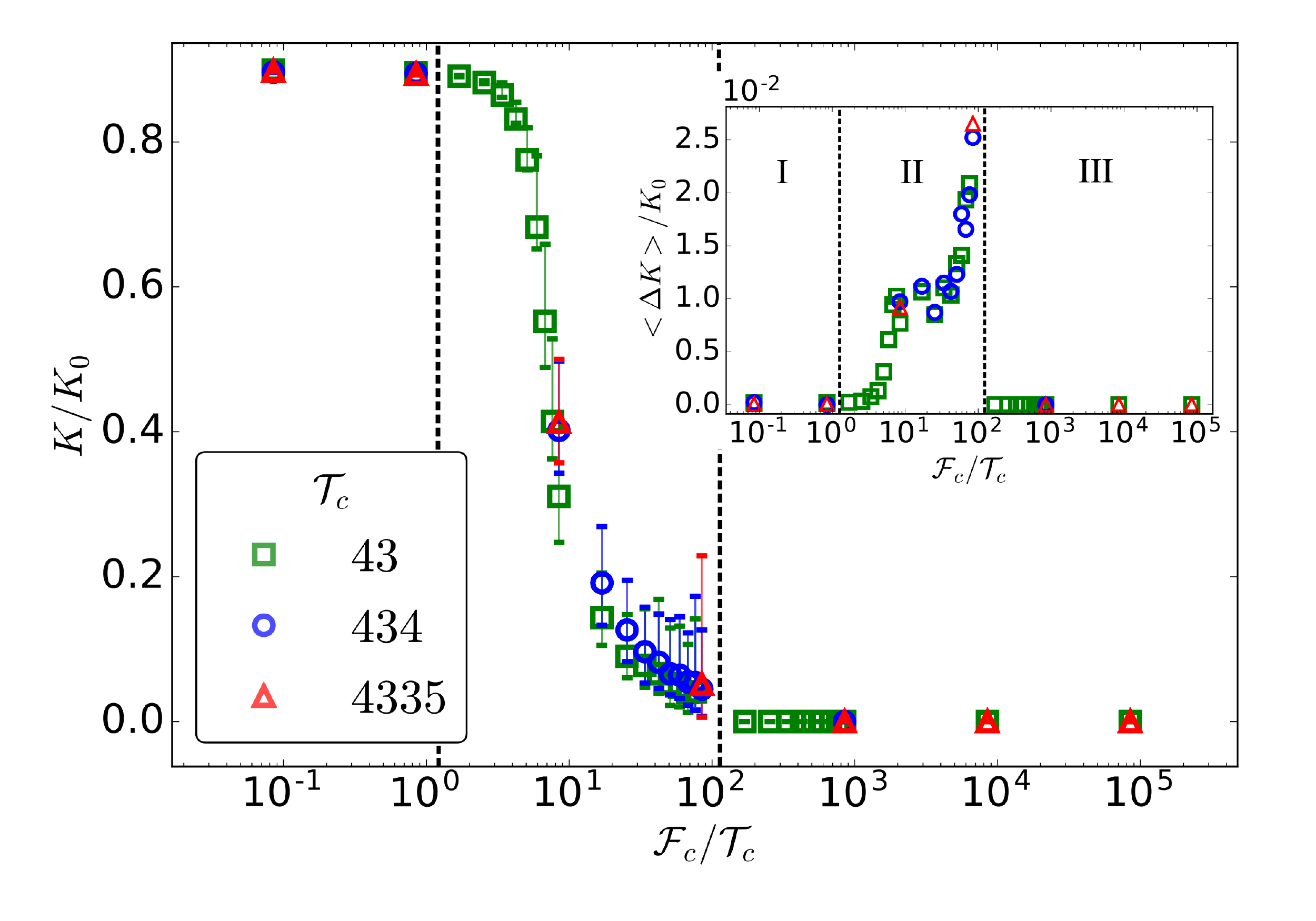}
\caption{Permeability change for constant pressure loss simulations. The plot shows the relative permeability after an initial deposition phase ($K_0$ is the permeability of the initial porous matrix), the error bars show the minimum/maximum permeability measured after the mean permeability has converged (there are still fluctuations due to bursts). For low $\mathcal{F}_c$ there is only slight deposition, as flow obstructing deposited matter gets immediately re-suspended. For high $\mathcal{F}_c$ the porous medium completely clogs and the permeability goes to zero. Within an intermediate range the permeability fluctuates around a finite value and rapid pressure jumps alternate with slow pressure increase. The inset shows the average relative jump in permeability. For the open phase (I) and the clogging phase (III) no jumps are observed, whereas for the transient phase (II) the relative jumps increase in size for higher $\mathcal{F}_c$.
}
\label{fig:perm}
\end{figure}

We also studied deposition and erosion for constant pressure loss $P=P_{\text{in}}-P_{\text{out}}$ imposed between in- and outlet, instead of constant flux. Here we define the characteristic speed as $u^* = K_0 P / \rho \nu L \epsilon$, where $K_0$ is the initial permeability and $L$ is the length of the porous matrix. As opposed to the case of constant flux, in this case the porous medium can completely clog and there are no fluctuations in the pressure loss. However sudden erosive events can still occur, because the local pressure difference can vary when matter is deposited. We set the threshold for shear erosion $\mathcal{T}_c$ such that the porous medium would completely clog, if channels could not reopen. For very high threshold $\mathcal{F}_c$ the local pressure difference is never high enough to lead to reopening (see Fig. \ref{fig:perm}), and for very low $\mathcal{F}_c$ any matter obstructing the flow will immediately detach. Here, we found again that for an intermediate range of $\mathcal{F}_c$ there is a intermittent regime where sudden erosive bursts are observed and the permeability fluctuates strongly. We found that the transition from the open phase to the clogged phase looks similar for different $\mathcal{T}_c$, and when we rescale $\mathcal{F}_c$ as follows:
\begin{equation}
\mathcal{F}_c \to \frac{\mathcal{F}_c} { \mathcal{T}_c } \propto \frac{1}{P} \frac{ \sigma_c }{ \tau _c },
\end{equation}
the measured permeabilities collapse (see Fig. \ref{fig:perm}), showing that the transition is only dependent on the ratio of the thresholds of the two erosive mechanisms. Sampaio et al. \cite{PhysRevLett.117.275702} developed an analogous model for an electric system, where a network of fuse-antifuse devices under a constant potential drop exhibit different phases: metallic, intermittent and insulating. Hence even though the operating conditions differ widely from imposing constant flux, here also three distinct regimes are found where only the middle range features conductive jumps and hence bursts in current, as can be seen in the inset of figure \ref{fig:perm}.

Our study shows that it is the re-opening of channels induced by hydraulic pressure that leads to the erosive bursts found in deep filtration experiments. We found that depending on material strength and flow conditions erosive bursts only occur for a certain range of thresholds for pressure induced erosion. We were able to reproduce the power-law behavior of pressure loss jumps found experimentally when erosive bursts occur in porous media. Furthermore we found that the power-law found by Bianchi et al. \citep{filippo} is universal in the sense that the exponent does not depend on Reynolds number nor on fluid properties. For both constant flux and constant pressure loss we found three phases. The first phase appears where the strength of deposited matter against pressure induced erosion is very weak and any deposit obstructing flow is re-entrained immediately. For intermediate strength some pressure has to build up to erode deposits, in this phase we can observe erosive bursts. In a third phase the strength against pressure induced erosion is too high for it to be a relevant mechanism for erosion and shear erosion is predominant. Thus we can conclude that it is not sufficient to only consider shear induced erosion in the study of erosion in porous media, but one has also to consider re-opening of channels induced by hydraulic pressure to be able to reproduce the behavior of erosive bursts and jumps in permeability in porous media.

Our findings help to better understand the erosive mechanisms in deep filtration, sudden sand production in oil wells and the evolution of erosion in dams and dikes. A future endeavor would be to construct experiments that allow to identify the strength of actual materials against shear and hydraulic pressure induced erosion and to characterize porous media for erosive behavior.

\begin{acknowledgments}
We thank F. Bianchi for providing the experimental data and for the valuable discussions. We acknowledge financial support from the European Research Council (ERC) Advanced Grant 319968-FlowCCS.
\end{acknowledgments}

\appendix
\section{Supplemental Material}
\label{sec:app}

\section*{Calculating the hydraulic pressure gradient}

To calculate the hydraulic pressure acting on a surface the momentum flux tensor $\pi$ needs to be calculated. In the lattice Boltzmann method (LBM) this can be done by using the distribution functions $f_i$ and the equilibrium distribution functions $f_i^{eq}$:

\begin{equation}
	\pi^{\alpha\beta} = \sum\limits_i \bigg[ \frac{1}{2\tau} f_i^{eq} + \bigg(1-\frac{1}{2\tau}\bigg)f_i \bigg] c_i^\alpha c_i^\beta ,
\end{equation}
where $\vec{c}_i$ are the lattice speed vectors of the LBM and $\tau$ is the relaxation time.

The normal vector $\hat{n}$ to the surface is given by the color gradient of the mass field:

\begin{equation}
	\hat{n}(\vec{x}) := -\frac{\sum\limits_{i} m(\vec{x}+\vec{c}_i)\cdot \vec{c}_i}
						{\sum\limits_{i} m(\vec{x}+\vec{c}_i)\cdot ||\vec{c}_i||_2}. 
\end{equation}

The solid mass field $m$ is a scalar field ($0 \leq m \leq 1$) specifying the fraction of a cell that is solid matter. When depositing or eroding matter ($m(\vec{x},t+\delta t) = m(\vec{x},t) \pm\delta m$) the exchanged matter $\delta m$ is removed respectively added to the concentration field ($C(\vec{x},t+\delta t) = C(\vec{x},t) \mp \delta m$). Note that $C$ is a volume concentration, giving the volume of solute divided by the total volume. See reference \cite{PhysRevE.95.013110} for the description of the algorithm. Thus while suspended particles are approximated using the convection-diffusion equation and hence point-like, they still occupy a volume, the volume of deposit is equal to 
\begin{equation}
V_{\text{deposit}} = m (\vec{x},t) \cdot \delta x^3 .
\end{equation}

 The hydraulic pressure is then calculated by summing up over all elements:

\begin{equation}
	P = \sum\limits_\alpha \sum\limits_\beta \pi^{\alpha\beta} n_\alpha n_\beta .
\end{equation}

This hydraulic pressure is measured on the surface of deposits, after which the maximum occurring pressure gradient through the deposit is calculated as follows:

\begin{equation}
	\nabla P_{a} = \underset{b}\max \frac{|P_a-P_b|}{||\vec{x}_a - \vec{x}_b||},
\end{equation}
where $a$ and $b$ are two points on opposite sides of a deposit. The calculated maximum pressure gradient $\nabla P$ is then compared with the threshold for hydraulic pressure induced erosion $\sigma_c$.

\section*{Jump size distributions for different Reynolds Numbers}

\begin{figure}
\includegraphics[width=\columnwidth]{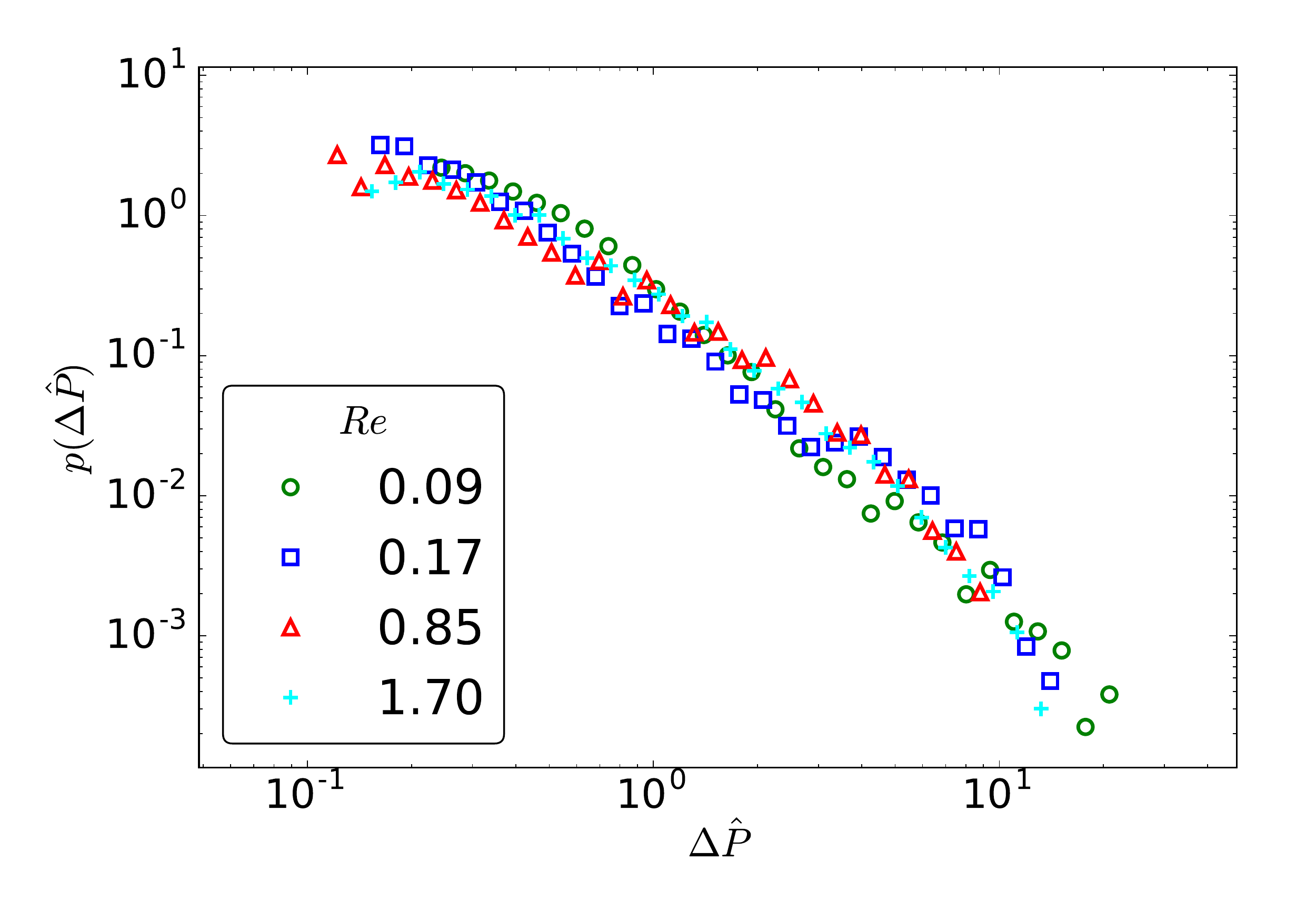}
\caption{Histogram of pressure loss jumps for different Reynolds numbers. The slope for the combined measurements is $\alpha = 2\pm 0.1$. 
}
\label{fig:rey}
\end{figure}

When increasing the magnitude of the imposed flux, the porous matrix clogs faster and also the pressure loss increases faster. However we found that even though the size of the pressure loss jumps changes, the power-law behavior still persists (see Fig. \ref{fig:rey}). 

\section*{Effect of Schmidt Number on erosive bursts}
\begin{figure}
\includegraphics[width=\columnwidth]{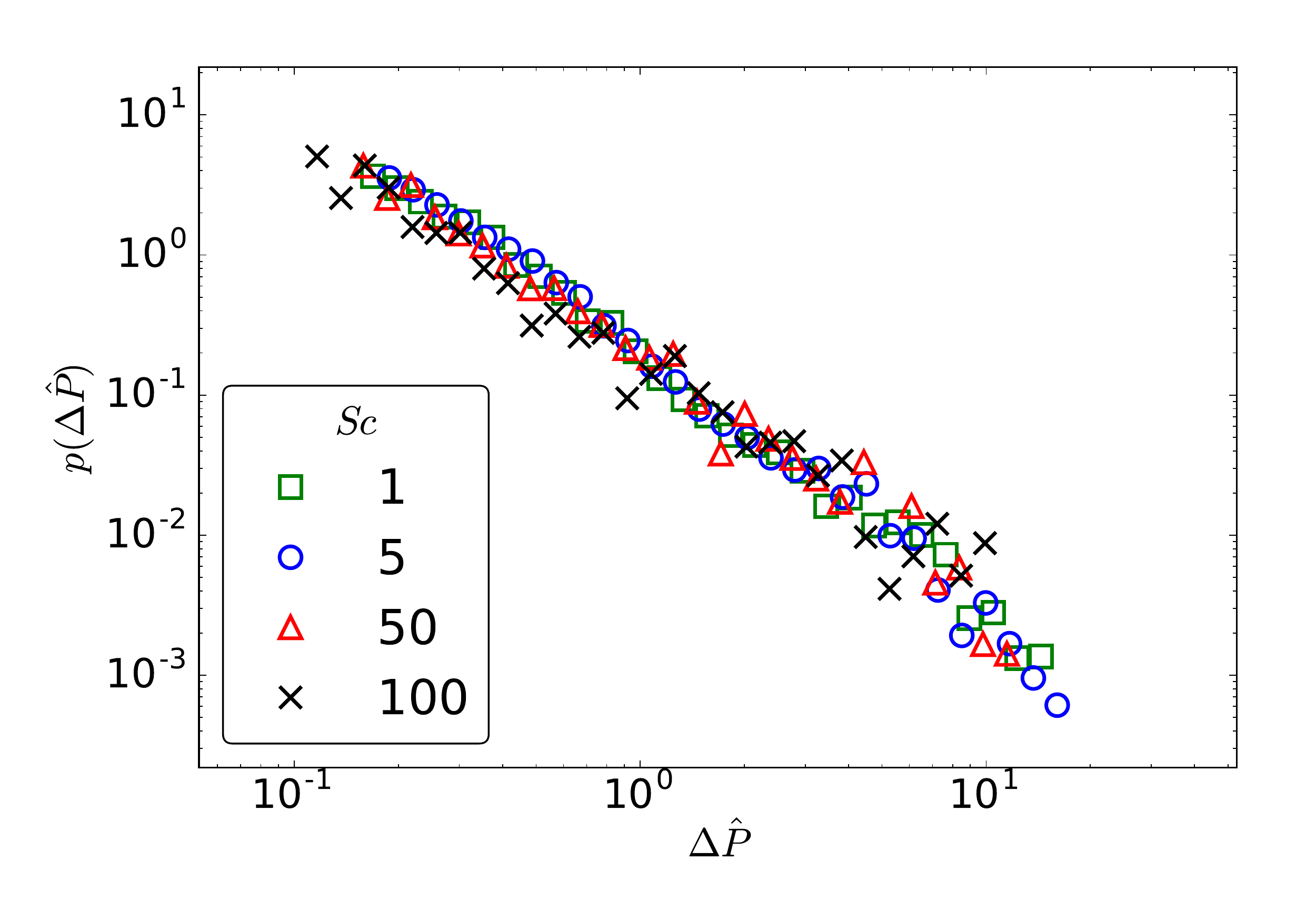}
\caption{Jump size distributions for different Schmidt numbers $Sc$.
}
\label{fig:schmidt}
\end{figure}

We found that increasing the Schmidt number leads to faster clogging of the porous medium and hence a faster increase of the pressure loss. However the overall behavior is still the same, the pressure loss jumps due to erosive bursts and the size of these jumps follow a power-law (see Fig. \ref{fig:schmidt}). The average jump size increases with increasing Schmidt number.

\section*{Boundary conditions reproducing experimental operating conditions}
In order to obtain the three phases observed, a more precise modeling of the experiment was required. For this purpose we imposed a constant flux until a critical pressure drop $P_{\rm max}$, was reached (representing the maximum pressure handled by the pump), and after that, this pressure drop was kept constant. In our simulations, we used $P_{\rm max} = 1.1 P_0$.

\section*{Frequency of erosive bursts as function of concentration}
\begin{figure}[h]
\includegraphics[width=\columnwidth]{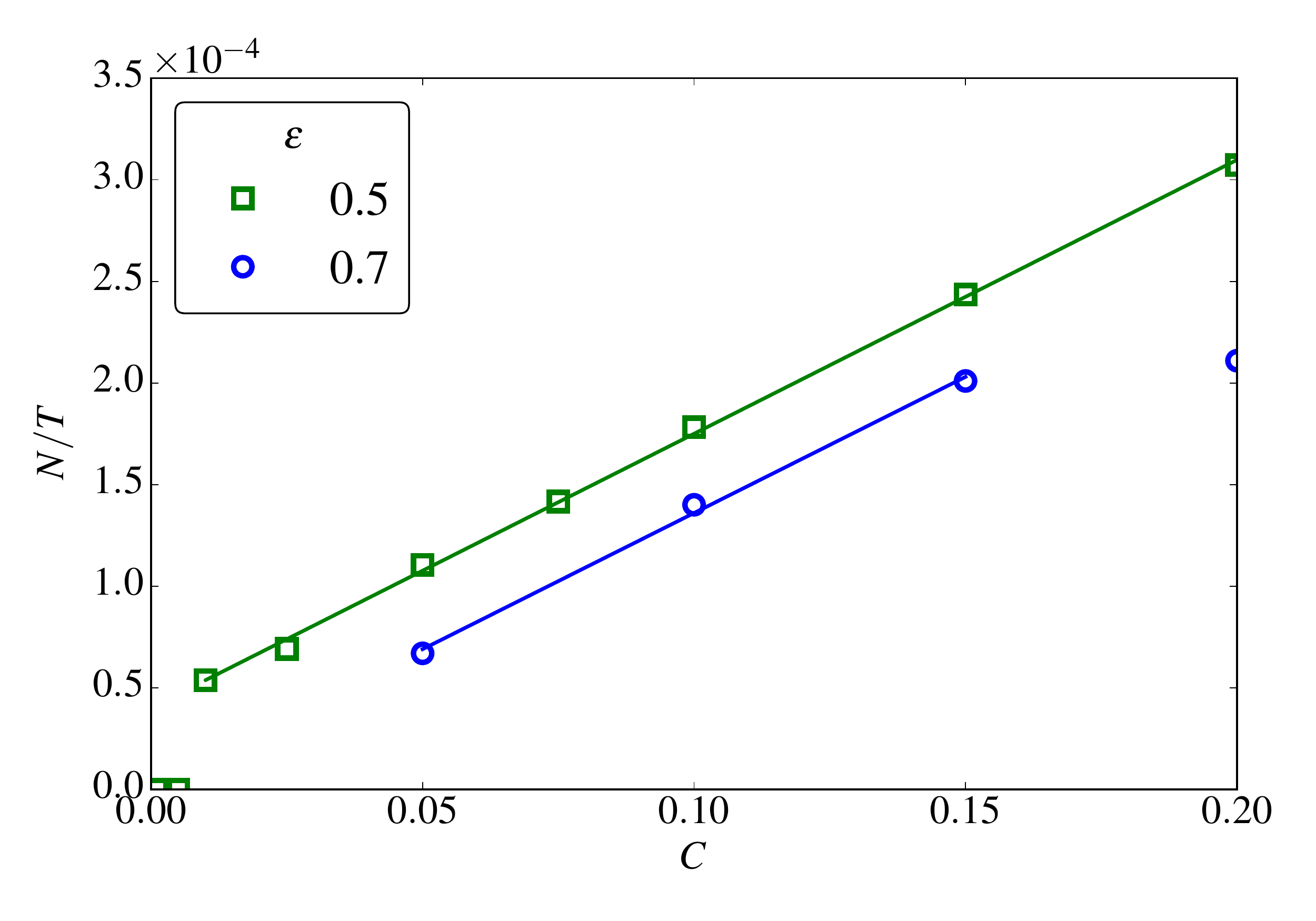}
\caption{The frequency of jumps ($N$ number of jumps, $T$ time from first to last jump) is plotted versus the concentration $C$ for two different porosities $\epsilon$.
}
\label{fig:jumpfreq}
\end{figure}
When the minimum concentration to allow erosive bursts is reached the frequency of jumps increases linearly (see Fig. \ref{fig:jumpfreq}) with the concentration. For very high concentrations ($C \sim 20\%$) we found deviations from this linear behavior for $\epsilon=0.7$.

\bibliographystyle{ieeetr}
\bibliography{citations.bib}
\end{document}